\documentclass[useAMS,usenatbib]{mn2e}

\usepackage{amsmath}
\usepackage{amssymb}
\usepackage[retainorgcmds]{IEEEtrantools}
\usepackage{graphicx}
\usepackage{pdftricks}
\begin{psinputs}
    \usepackage{pstricks}
    \usepackage{multido}
\end{psinputs}
\usepackage{color}
\usepackage[T1]{fontenc}

\newcommand{\uBH}{\mathrm{BH}}
\newcommand{\uEdd}{\mathrm{Edd}}
\newcommand{\uEM}{\mathrm{BZ}}
\newcommand{\uF}{\mathrm{F}}

\newcommand{\uH}{\mathrm{H}}

\newcommand{\uLT}{\mathrm{LT}}

\newcommand{\uMHD}{\mathrm{MHD}}
\newcommand{\uT}{\mathrm{T}}
\newcommand{\uacc}{\mathrm{acc}}
\newcommand{\uc}{\mathrm{c}}

\newcommand{\ud}{\mathrm{d}}
\newcommand{\udisc}{\mathrm{disc}}

\newcommand{\uff}{\mathrm{ff}}

\newcommand{\ufp}{\mathrm{fp}}
\newcommand{\ug}{\mathrm{g}}
\newcommand{\ugas}{\mathrm{gas}}

\newcommand{\uin}{\mathrm{in}}

\newcommand{\um}{\mathrm{m}}
\newcommand{\umag}{\mathrm{mag}}

\newcommand{\uout}{\mathrm{out}}
\newcommand{\up}{\mathrm{p}}

\newcommand{\urad}{\mathrm{rad}}
\newcommand{\us}{\mathrm{s}}

\newcommand{\utot}{\mathrm{tot}}

\newcommand{\sgra}{Sgr~A$^*$}
\newcommand{\grs}{GRS~1915+105}

\newcommand{\aap}{A\&A}

\newcommand{\apj}{ApJ}
\newcommand{\apjl}{ApJ}

\newcommand{\araa}{ARA\&A}

\newcommand{\mnras}{MNRAS}

\newcommand{\pasj}{PASJ}

\newcommand{\ssr}{Space Science Reviews}

\title[Electromagnetic vs. Lense-Thirring alignment]{Electromagnetic vs. Lense-Thirring alignment of black hole accretion discs}
\author[P. Polko and J. C. McKinney]{Peter Polko$^1$\thanks{E-mail: polko@umd.edu (PP); jcm@umd.edu (JCM)} and Jonathan C. McKinney$^1$\\
$^1$University of Maryland at College Park, Dept. of Physics, Joint Space-Science Institute,\\
\;\;3114 Physical Sciences Complex, College Park, MD 20742, USA \\
}{
}

\begin{document}

\pagerange{\pageref{firstpage}--\pageref{lastpage}}

\maketitle

\label{firstpage}

\begin{abstract}

Accretion discs and black holes (BHs) have angular momenta that are
generally misaligned with respect to each other, which can lead to
warps in the discs and bends in any jets produced. We consider a disc
that is misaligned at large radii and torqued by Lense-Thirring (LT)
precession and a Blandford-Znajek (BZ) jet torque.  We consider a
variety of disc states that include radiatively inefficient thick
discs, radiatively efficient thin discs, and super-Eddington accretion
discs.  The magnetic field strength of the BZ jet is chosen as either
from standard equipartition arguments or from magnetically arrested
disc (MAD) simulations.  We show that standard thin accretion discs
can reach spin-disc alignment out to large radii long before LT would
play a role, as caused by the slow infall time that gives even a weak
BZ jet time to align the disc. We show that geometrically thick
radiatively inefficient discs and super-Eddington discs in the MAD
state reach disc-spin alignment near the black hole when density
profiles are shallow as in magnetohydrodynamical simulations, while
the BZ jet aligns discs with steep density profiles (as in
advection-dominated accretion flows) with the BH spin out to larger
radii.  Our results imply that the BZ jet torque should affect the
cosmological evolution of BH spin magnitude and direction, BH spin
measurements in active galactic nuclei and X-ray binaries, and the
interpretations for Event Horizon Telescope observations of discs or
jets in strong-field gravity regimes.

\end{abstract}


\section{Introduction}
\label{sec:introduction}

Jets are ubiquitous in the Universe.  While their formation mechanisms
are still unclear, the required components appear to be a magnetic
field, rotation, and accretion. In the case of black holes (BHs), the
accretion is usually expected to take the form of a Keplerian disc
with an angular momentum axis defined by the rotation of the disc
material. The angular momentum of the source of disc matter, however,
is capable of being arbitrarily misaligned with respect to the BH spin
axis.  Because BH jets are so prevalent, either the initial rotation
of the matter is not important to jet formation, or the matter is
torqued in some way to align with the angular momentum vector of the
black hole. The gravitational potential of a rotating black hole is
not spherically symmetric, so a different orientation of the disc
could also have a significant effect on emergent radiation, affecting
BH spin measurements \citep{2011MNRAS.414.1183K,2014SSRv..183..295M},
jet orientation \citep{2013Sci...339...49M}, and timing features in
light curves \citep{2009MNRAS.397L.101I,2013MNRAS.432.2252D}.  Hence,
determining whether (and to what degree) discs align by the action of
BH spin is important.

There are a few processes through which material is torqued: 1) the
exchange of angular momentum between neighbouring rings of material, a
process called the accretion torque; 2) the Lense-Thirring effect, which is due
to the dragging of spacetime by a rotating black hole combined with
viscous forces \citep{1918PhyZ...19..156L}; and 3) a
\citet{1977MNRAS.179..433B} (hereafter BZ) jet that is aligned with
the rotation of the black hole near the black hole, but misaligned at
larger radii by it trying to push on a misaligned disc
\citep{2013Sci...339...49M}.

A study into magnetic alignment, corresponding to the third process,
was performed by \citet{1977A&A....58..175K} (hereafter KL, see also
\citealt{2005MNRAS.363...49K}), who concluded that either the inflow
timescale was shorter than the alignment timescale, or the accretion
rate was so low that the source could not be observed in the first
place. Their negative result hinged on the assumptions that 1) the
maximum magnetic field strength was limited by weak turbulent eddies
in the disc; 2) the disc is rather small in radial extent with a
relatively short inflow time; and 3) the disc followed the
Shakura-Sunyaev density profile generalised by
\citet{1973blho.conf..343N} (hereafter NT).

Since KL's study on magnetic alignment, several disc models have been
proposed, such as geometrically thick radiatively inefficient
accretion flows (RIAFs) as the advection dominated accretion flow
(ADAF) model \citep{1994ApJ...428L..13N,1995ApJ...444..231N}.  The
ADAF model has a steeper density profile than the NT solution, so a BH
jet can more easily torque material at larger radii.  The nature of
the magnetic field can also be quite different, such as in the
magnetically arrested disc (MAD) model \citep{2003PASJ...55L..69N}. In
the MAD model, the magnetic field builds up near the black hole, until
the magnetic tension is high enough that it balances inward forces due
to the massive accretion flow. The magnetic field strengths attained
are higher than previously considered, which means that magnetic
alignment may yet actually be an efficient method of aligning the
disc.  Lastly, astrophysical discs can extend to tens of thousands of
gravitational radii, where the inflow time becomes quite long and
allows for the BZ jet torque to cumulate.

In this paper, we investigate how a disc aligns due to the
Lense-Thirring and BZ jet torques by computing the disc shape from
angular momentum conservation.  We will demonstrate that disc
alignment can occur for a variety of accretion rates, BH masses, and
magnetic field strengths.

In Section \ref{sec:method} we will describe the equations we use to
compare the different processes. In section \ref{sec:discsystems} we will discuss the different accretion systems and disc models we will use. In Section \ref{sec:results} we will
show the alignment of the different disc models and for a variety of
systems such as \sgra{} and \grs{}. In Section \ref{sec:discussion} we
will discuss the results, and in section \ref{sec:conclude} we
will present our conclusions.

\section{Method}
\label{sec:method}

In this section, we will obtain our equations of motion, give a
description of the different scalings, and derive the external torques.

\subsection{Equations of Motion}

We start with the general relativistic magnetohydrodynamic (GRMHD)
equations of motion, given by energy-momentum conservation:
\begin{equation}
\nabla_\mu (T^\mu_\nu) = G_\nu,
\end{equation}
for an ideal stress-energy tensor $T^\mu_\nu = (\rho + p_\ug + u_\ug + b^2) u^\mu
u_\nu + \delta^\mu_\nu p_\utot - b^\mu b_\nu$ (given in terms of a
coordinate basis), where we assume the induction equation controls the
evolution of the magnetic field.  The source term $G_\nu$ includes any
disc cooling or external torques on the gas.  After the usual
height-integration and sticking to an orthonormal vector basis, one
obtains for the $\phi$ equation of motion:
\begin{equation}\label{tauacc}
\frac{1}{r} \partial_r (r^2 \Sigma v^r v^\phi) = \tau_\uacc,
\end{equation}
where $\Sigma = \int_\theta \rho r \,\ud\theta = 2 \rho H$ and
$\tau_\uacc$ is the accretion torque per unit area that includes all
magnetic and viscous terms that drive accretion and angular momentum
transport \citep{2002apa..book.....F}.  The height-integrated angular
momentum per unit area is given by $\tilde{L} = r \Sigma v^\phi$, so
that one can write the $\phi$ equation of motion as just
\begin{equation}
\frac{1}{r} \partial_r (r v^r \tilde{L}) = \tau_\uacc.
\end{equation}
The equation in vector form for all components that is valid in the small-angle approximation is
\begin{equation}
\frac{1}{r} \partial_r (r v^r \vec{\tilde{L}}) = \vec{\tau},
\label{vectau}
\end{equation}
where $\tau$ is the total vector torque per unit area that includes
the accretion torque and any other external torques
(i.e. Lense-Thirring, Blandford-Znajek jet, etc.).

\begin{figure}
\begin{center}
\includegraphics[width = 0.46 \textwidth]{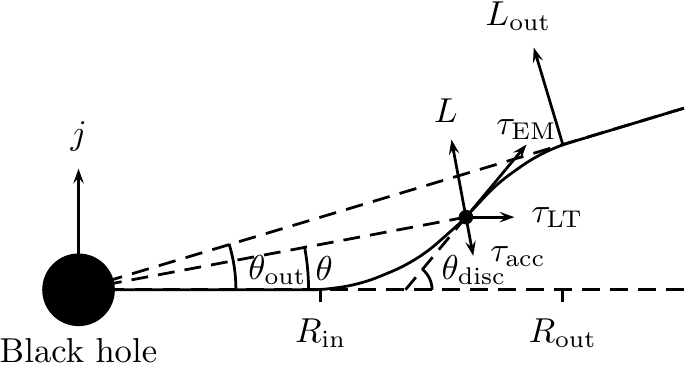}
\end{center}
\caption{Schematic of the model used showing a slice through one side
  of the disc that has the most geometric misalignment with respect to
  the black hole.  The disc is assumed to be non-axisymmetrically
  warped with no substantial twist. The disc starts out at an outer
  radius with an outer position angle $\theta_\uout$ compared to the
  plane at $\theta = 0$ of the black hole spin vector $j$. The accretion
  disc has angular momentum $L_\uout$ and accretion is mediated by the
  accretion torque ($\tau_\uacc$), which is always parallel to the
  angular momentum vector. At some radius ($R_\uout$), the
  Blandford-Znajek ($\tau_\uEM$) and Lense-Thirring torque
  ($\tau_\uLT$) start changing the direction of the angular momentum
  vector, and thereby the shape of the disc. The electromagnetic
  torque lies along the surface of the disc, while the Lense-Thirring
  torque points along the black hole spin vector plane for maximal
  alignment by forces driven by viscous action on Lense-Thirring
  precession torques. We calculate the magnitude and direction of the
  angular momentum ($L$), and hence the position angle ($\theta$) and
  the disc angle ($\theta_\udisc$) of the disc (solid line) from the
  torques. If the disc has aligned to within a certain small angle
  relative to the initial angle $\theta_\uout$, then we consider
  alignment effectively complete at that radius ($R_\uin$).}
\label{fig:schematic}
\end{figure}

\subsection{Setup}

Fig.~\ref{fig:schematic} shows the general setup of our problem of a
warped disc produced by torques that act to align an outer disc that
is misaligned with respect to the BH spin axis.  We assume the
external torques all act to re-align the disc and do not cause
twisting, precession, or breaking.  So, the angular momentum vector of
the disc reorients towards the BH spin axis without any twist.  Then,
there are only two angular momentum components to consider and so only
two equations of motion to consider (i.e. the vector parallel to the
BH spin axis and that perpendicular to the spin axis).

We assume small misalignments, so that the vector component for the
misaligned angular momentum of $|\tilde{L}| \sin(\theta)$ is the required
lowest-order term, meaning any dependence upon $\theta$ within the
full tilted disc solution for $\Sigma$, $v^r$, or $v^\phi$ can be
neglected as being a higher-order correction.  Here $\theta$ is the
angle away from the equatorial plane, and toward the $z$ axis is
positive $\theta$.  Similarly, for the aligned angular momentum of
$|\tilde{L}| \cos(\theta)$, the perturbed disc solution would only be
higher-order in $\theta$ and can be neglected for small tilts.  This
means that we can assume the radial profiles of $\Sigma$, $v^r$, and
$v^\phi$ correspond to the original (e.g.) NT, ADAF, or untilted
simulation profiles.  This is why we do not have to consider the
rest-mass or energy equations of motion.

The assumption of small tilt angles also means the radial direction
and path length are approximately equal to the true disc direction and
true disc path length, so we do not need to distinguish (say)
$\vec{r}$ from $\vec{l}$, which would be the actual vector position of
the disc.  This means $v^r$ is a sufficiently accurate velocity along
the disc path.

We calculate the magnitude of the angular momentum per unit area
$\tilde{L}$ at some outer radius $R_\uout$ at angle $\theta_\uout$ and
decompose it into an aligned component
$L_\parallel = |\tilde{L}| \cos(\theta)$, parallel to the spin vector of
the black hole, and a misaligned component
$L_\perp = |\tilde{L}| \sin(\theta)$, pointing perpendicular to this
direction.  So, we have starting values $L_{\uout \parallel} =
|\tilde{L}| \cos(\theta_\uout)$ and $L_{\uout \perp} = |\tilde{L}|
\sin(\theta_\uout)$ at an outer radius $R_\uout$ and outer disc
position angle $\theta_\uout$, where the starting values use
$|\tilde{L}|$ from the untilted disc model.  We use Eq.~(\ref{vectau})
to decompose the angular momenta into parallel and perpendicular
components, giving:
\begin{eqnarray}\label{vectauperppar}
\tau_{\utot \parallel} &=& \frac{1}{r} \partial_r (r v^r \tilde{L}_{\parallel}),\\
\tau_{\utot \perp} &=& \frac{1}{r} \partial_r (r v^r \tilde{L}_{\perp}).
\end{eqnarray}

We solve these torque equations discretely for new values of
$\tilde{L}_\parallel$ and $\tilde{L}_\perp$ (giving $\theta$) for a
radial step size $\Delta R \propto -R$, while the new disc angle
$\theta_\udisc$ is obtained from the difference in positions of the
disc:
\begin{IEEEeqnarray}{rCl}
R_{n+1} & = & R_n + \Delta R,\\
\theta_{n+1} & = & \arctan(-L_\perp / L_\parallel),\\
\theta_{\udisc, n+1} & =& \arctan \left[ \frac{R_{n+1} \sin(\theta_{n+1}) - R_n \sin(\theta_n)}{R_{n+1} \cos(\theta_{n+1}) - R_n \cos(\theta_n)} \right],
\label{thetanp1}
\end{IEEEeqnarray}
which after the first step no longer requires information from the
untilted disc model for $\Sigma$ or $v_\phi$ -- only $v^r$ is still
required.  In our small-angle approximation, the parallel torque
equation just becomes an equation that ensures the accretion torque
drives the correct behavior for the total angular momentum.  The
perpendicular torque equation contains all information about the
actual angle, so in principle it can be solved for $\theta$ directly
by plugging in the approximation that $\tilde{L}_{\perp} \approx
|\tilde{L}| \sin(\theta)$ for all radii using the original untilted
disc model value of $|\tilde{L}|$.  We choose a slightly different but
same-order accuracy approach, by evolving $\tilde{L}_\parallel$ and
$\tilde{L}_\perp$ independently from each torque equation to obtain
$\theta$ as in Eq.~(\ref{thetanp1}).  This results in a total
$\tilde{L}$ not identical with the untilted disc model value, but that
is consistent with our small-angle approximation and the higher-order
error becomes negligible for small $\theta$.  We use at least ten
steps per decade in radius, and to ensure accuracy we half the step
size if the change in $\theta$ for any step is more than $\theta_\uout
/ 10$.  We also check that our choice of initial angle leads to an
$\tilde{L}$ similar to the untilted disc model value as must be true
if we have chosen a sufficiently small angle so our approximations are
valid.

\subsection{Mass and Distance Scaling}

Because we have systems with a wide range of masses, we will use the
mass of the black hole in solar masses ($m = M / M_\odot$) as our mass
scale, the corresponding gravitational radius as our length scale
($\bar{r} = r / r_\ug$, with $r_\ug = G M / c^2$), and the light crossing
time as our time scale ($t_\ug = G M / c^3$), with $G$ the gravitational
constant, and $c$ the speed of light. We denote with a bar every
variable that follows this scaling ($r \rightarrow \bar{r}$), and use
small letters, or a different symbol for every variable that has an
alternate scaling (i.e. total angular momentum of the BH as $J
\rightarrow j$ that is the dimensionless BH spin, and magnetic flux
$\Phi \rightarrow \Upsilon$ as the dimensionless magnetic flux). We
scale the accretion rate to the Eddington accretion rate for a one
solar mass black hole,
\begin{equation}
\dot{m} \equiv \dot{M} / \dot{M}_\uEdd = \dot{M}_\uBH / (\dot{M}_{\uEdd \odot} m) \bar{r}^s ,
\end{equation}
where $\dot{M}_\uBH$ is the accretion rate at the horizon, $0 \leq s
\leq 1$ allowing for a wind if $s > 0$, and where
\begin{equation}
\dot{M}_{\uEdd \odot} = \frac{4 \pi G M_\odot m_\up}{\eta c
  \sigma_\uT} = \frac{1.40 \times 10^{17}}{\eta}~\ug~\us^{-1},
\end{equation}
where $m_\up$ is the mass of a proton, $\eta$ is the
arbitrarily-chosen nominal accretion efficiency of $\eta=0.1$, and
$\sigma_\uT$ is the Thomson cross section. We normalise the spin to
the mass of the black hole, $j = \frac{c}{G M_\odot^2} m^{-2} J$. The
dimensionless surface density is given by:
\begin{equation}
\bar{\Sigma} = \frac{G^2 M_\odot}{c^4} m \Sigma,
\end{equation}
and the dimensionless torque per unit area is:
\begin{IEEEeqnarray}{rCl}
\bar{\tau} & = & \frac{G^2 M_\odot}{c^6} m \tau = 1.22 \times 10^{-44} m \tau.
\end{IEEEeqnarray}

\subsection{Lense-Thirring Torque}

The dragging of spacetime by a rotating black hole aligns material via
the Lense-Thirring (LT) effect \citep{1918PhyZ...19..156L}.  The LT
effect inevitably leads to precession of the disc, but neighbouring
rings can exchange angular momentum through viscosity, allowing
alignment of the disc. Although this effect is only efficient very
close to the black hole, the torque can also propagate outward via
viscous forces, a process known as the Bardeen-Petterson effect
\citep{1975ApJ...195L..65B}, which is commonly invoked as a method to
align accretion discs with the spin of the black hole.

We assume the LT torque does not cause substantial twisting, so that
twist angles are much smaller than the assumed small tilt angles.  We
also assume the disc does not break, as can occur for large tilt
angles \citep{2012MNRAS.421.1201N}.  Lastly, we assume the disc does
not precess, but we check our solutions for when precession is likely
to occur (i.e. $\alpha > H / R$ for viscosity parameter $\alpha$ and
height-to-radius ratio of $H / R$).  In such cases when LT dominates
alignment and $\alpha \gtrsim H / R$, we check if the BZ jet torque would
have led to alignment anyway without the LT torque.  We neglect
apsidal precession, which could lead to additional effects.

The LT precession torque is given by
\begin{equation}
\tau_\uLT \approx |\boldsymbol{\Omega}_\uLT \times
\boldsymbol{\tilde{L}}| \approx \Omega_\uLT \tilde{L} \sin(\theta),
\end{equation}
where $\Omega_\uLT = 2 j / r^3$ describes the frame-dragging.
For a general disc model, we assume a scaled Keplerian motion, so that the
rotational velocity per unit speed of light is given by
\begin{equation}
\beta_\phi = \beta_0 \bar{r}^{-1/2} = v_\phi / c.
\end{equation}
So the dimensionless form of the LT torque is given by
\begin{equation}
\bar{\tau}_\uLT \approx 2 j \bar{\Sigma} \beta_0 \sin(\theta)
\bar{r}^{-5/2}.
\end{equation}

Nominally some viscous process acts to redistribute the precession
torque that could cause alignment in some cases.  We assume the LT
precession torque is re-directed by viscous interaction between
differential precession as to maximally align the disc.  This means
that the LT torque magnitude is taken as the precession torque, but
the direction is taken as that which maximally aligns the disc:
\begin{eqnarray}
\tau_{\uLT,\parallel} & \approx & 0,\\
\tau_{\uLT,\perp} & \approx & \Omega_\uLT \tilde{L} \sin(\theta).
\end{eqnarray}
This is generally not the case, but the final solution does not depend
upon the LT torque direction being exactly along the perpendicular
direction, and instead it could have been oriented perpendicular to
the accretion torque or oriented along the disc like the
electromagnetic torque.  To lowest order in $\theta$, these all give
the same resulting torque and behavior for $\tilde{L}$.

Our goal is to conservatively estimate whether the competing BZ jet
torque can dominate any effect by LT.  By choosing the maximal LT
alignment effect, we give the LT torque the best chance to dominate
the BZ jet torque physics.  If our results show this conservative
choice still leads to the BZ jet dominating the LT alignment, then our
result of BZ dominance is robust to any weaker LT effect.  Given the
current significant uncertainty in how (or whether) the LT and other
general relativistic effects lead to alignment, we take this
simplifying approach in order to make restrictive statements about
strength or physics involved in LT leading to alignment.  If LT ends
up causing less alignment, precession, or breaking, this only weakens
the role of LT in aligning the disc and allows the BZ jet to more
generally dominate.

\subsection{Blandford-Znajek Jet Torque}

If the electromagnetic (EM) jet is aligned with the rotation of the
black hole, it can push on one side of a misaligned disc
\citep{2013Sci...339...49M}. The torque described by KL was derived
assuming vacuum conditions, so there is no spin-down or torque when
the field is aligned with the black hole spin axis.  However,
\citet{1977MNRAS.179..433B} showed that even if the field is aligned
with the BH spin axis, the BH will release angular momentum.  In
general, any orientation of the magnetic field will produce torques in
the surrounding disc when the jet satisfies the conditions for
force-free or magnetohydrodynamics (MHD).  A BZ jet contains a
constant magnetic flux and rotates at a constant field line rotation
frequency along field lines (and individual field lines rotate at
similar frequencies to order unity).  For small spins, the spin axis
does not necessarily control the jet direction even near the BH, while
for larger spins the jet direction near the BH is dominated by the
spin axis \citep{2013Sci...339...49M}.  We assume the regime where the
spin magnitude is large enough to have the jet aligned with the BH
near the BH, although arbitrary angles applicable to small spins could
be considered.

The BZ jet torque on the disc is due to the magnetic field draping the
surface of the disc and pushing into the disc surface with a magnetic
pressure. KL found that the torque along the azimuthal direction (into
the plane in Fig.~\ref{fig:schematic}) vanishes, so there is no
precessing torque.  So the BZ jet torque has components
\begin{eqnarray}
G_{\uEM\parallel} & = & |G_{\uEM}| \sin(\theta_\udisc),\\
G_{\uEM\perp} & = & |G_{\uEM}| \cos(\theta_\udisc),\\
G_{\uEM-y} & = & 0,
\end{eqnarray}
where KL found $|G_\uEM| \propto \sin(\theta)$.  One could set
$\theta_\udisc \to \theta$ and our results are not substantially
changed.  We obtain the net torque per unit area $|\tau_{\uEM}| =
|G_{\uEM}| / A$ for some area $A$ from $|G_{\uEM}| / A = |\vec{r} \times
\Delta \vec{F}| / A = r (|\Delta F| / A) \cos(\theta - \theta_\udisc)$.  The
net force vanishes for zero tilt due to the jet pushing on the disc
from above and below equally, while when the tilt is finite then the
jet pushes down more on one side than the other.  So $\Delta F \approx
|F| \sin(\theta)$ for a one-sided force $|F|$, which gives
\begin{equation}
|G_\uEM| \approx r (|F| / A) \sin(\theta) \cos(\theta - \theta_\udisc) .
\end{equation}
As the field rotates, the field could lag in shape and not follow the
disc surface, which could introduce order unity changes to our force
difference $\Delta F$.  The one-sided force per unit area $|F|/A$ is
due to the magnetic pressure $p_b$ of the jet pushing on the disc
orthogonal to the disc surface.  This gives that the net torque per
unit area of the jet pushing down on the disc is
\begin{eqnarray}
\tau_{\uEM \parallel} & \approx & r p_b \sin(\theta) \cos(\theta - \theta_\udisc) \sin(\theta_\udisc),\\
\tau_{\uEM \perp} & \approx & r p_b \sin(\theta) \cos(\theta - \theta_\udisc) \cos(\theta_\udisc),
\end{eqnarray}
which is consistent with KL's torque direction and magnitude in the
limit they considered where $\theta_\udisc \to \theta$. The magnetic
pressure is $p_b = b^2 / (8 \pi)$ (for $b$ in Gaussian units), $b^\mu =
\frac{1}{2} \epsilon^{\mu \nu \kappa \lambda} u_{\nu} F_{\lambda
  \kappa}$ is the magnetic field 4-vector, $\epsilon$ is the
Levi-Civita tensor, $u$ is the 4-velocity, and $F$ is the
electromagnetic tensor
\citep{2003ApJ...589..444G,2006MNRAS.367.1797M}. In flat space-time,
$b^2 = \frac{B^2 + (u \cdot B)^2}{\gamma^2}$.  At large radii for a
jet, $b^2 \approx \frac{B_\phi^2}{\gamma^2}$, because the field
becomes toroidal and the velocity becomes radial, while at small radii
$b^2\approx B_r^2$ with a small angular velocity dominating the radial
velocity.  So, $(u\cdot B) \approx 0$ is generally small compared to
the other term. The lab-frame magnetic field's radial component $B_r$
is determined by the magnetic flux lines given by
\begin{equation}
\Phi_G(\theta) = (1/2) \int_{\theta \phi} r^2 \sin(\theta) |B_r| \,\ud \theta \,\ud \phi,
\end{equation}
with radial magnetic field $B_r$ in Gaussian units, and where
$\Phi_G(\theta \approx \pi / 2) \approx 2 \pi r^{2 - \nu} B_r$ with jet
collimation parameter $0 \leq \nu \lesssim 1$ described by the field
shape
\begin{equation}
\Phi_G(\theta) \propto r^\nu [1 - \sin(\theta)],
\end{equation}
which is some constant $d$ for a field line of interest that closely
hugs the disc surface with $\nu \approx 0$.  The parameter $\nu$ could
be derived by pressure-matching the jet magnetic pressure with the
disc gas pressure for a given disc model
\citep{2007MNRAS.375..513M,2007MNRAS.375..531M}. In our case, we do
not model the role of gas pressure, so the disc-jet interaction takes
place as effectively having the magnetic field fill most of space for
small tilt disc angles and small disc curvatures as if $\nu \approx 0$.
We use a dimensionless magnetic flux passing through the BH horizon
(at $r \approx r_\uH$, the horizon radius) of
\begin{equation}
\Upsilon \approx 0.7 \Phi_G[r = r_\uH] / \sqrt{4 \pi r_\ug^2 \dot{M}_\uBH c}.
\end{equation}
For force-free jets, the radial field is approximately related to the
toroidal field by
\begin{equation}
\label{bphi}
B_\phi \approx B_r [r \Omega_\uF \cos(\theta)] / c,
\end{equation}
\citep{2007MNRAS.375..548N}, for field line angular frequency of
rotation $\omega_\uF = r_\ug \Omega_\uF / c \approx j / (4 r_\uH)$,
where the last approximation comes from the Blandford-Znajek split
monopole valid at high $j$
\citep{1977MNRAS.179..433B,2010ApJ...711...50T,2012MNRAS.423L..55T,2012JPhCS.372a2040T}.
Thick discs could restrict some energy outflow and lead to a more
rapid change in power as a function of $j$ \citep{2005ApJ...630L...5M}, but the
total power output is the same at high $j$.

For an MHD jet, the behaviour of $B_\phi$ can be derived from energy
conservation via the definition of $\mu$ that is the energy flux per
unit rest-mass flux related to the conversion of electromagnetic to
kinetic energy. This gives that
\begin{equation}
B_\phi \approx B_{\phi,\ufp} \left( \frac{\gamma-\mu}{\gamma_\ufp-\mu} \right) \left( \frac{r_\ufp \Omega_{\uF,\ufp}}{r \Omega_\uF} \right),
\end{equation}
so that for given foot point values ($B_{\phi,\ufp} \approx B_{r,\ufp} [r_\ufp \Omega_{\uF,\ufp} \cos(\theta_\ufp)] / c$,
$r_\ufp\approx r_\ug$, $\gamma_\ufp\approx 1$, $\Omega_{\uF,\ufp}
\approx \Omega_\uF$ for our $\nu \approx 0$ case), we need
$\gamma(r)$ to determine $B_\phi(r)$ \citep{2012MNRAS.419..573M}. This
MHD version of $B_\phi$ would not include any ram pressure or thermal
pressure that would act to push the disc into alignment if the jet
were aligned with the BH spin axis as it would be if the jet started
as electromagnetic.  The ram pressure is $\rho_0
\gamma(\gamma-1) \approx B_r \gamma/\phi$ for conserved quantity
$\phi \propto B_r / (\rho_0 \gamma)$.  The magnetic pressure dropping
faster than $1/r^2$ once MHD acceleration occurs is therefore
compensated by the ram pressure that still goes as $1 / r^2$ for
$B_r \propto r^{-2}$ with $\gamma$ going from $1 \to \mu$.  So we
generally use Eq.~(\ref{bphi}) for $B_\phi$ to account for the total
jet torque.

For the Lorentz factor's radial dependence, we use the results from
analytical calculations and simulations
\citep{2008MNRAS.388..551T,2009ApJ...699.1789T,2010NewA...15..749T,2012MNRAS.419..573M}. The
Lorentz factor is given by
\begin{equation}
\frac{1}{\gamma} \approx \frac{1}{\gamma_\uff} + \frac{1}{\mu},
\end{equation}
where $\gamma_\uff$ is the force-free Lorentz factor, and $\mu$ is the
total energy flux per unit mass flux (and the upper limit of
$\gamma$). Furthermore,
\begin{equation}
\frac{1}{\gamma_\uff^2} \approx \frac{1}{\gamma_1^2} + \frac{1}{\gamma_2^2},
\end{equation}
where
\begin{equation}
\gamma_1 \approx \sqrt{\gamma_0^2 + (\Omega_\uF R/c)^2 - (\Omega_{\uF,\ufp}
  R_\ufp/c)^2},
\end{equation}
describes the Lorentz factor first asymptotic regime due to the
toroidal winding of the field lines, with $\Omega_{\uF,\ufp}$ the
field rotation frequency, and $R_\ufp$ the radius at the footpoint of
the field line.  For curved field lines (i.e. $\nu > 0$), the second
asymptotic is given by
\begin{equation}
\gamma_{2,\nu} \approx \sqrt{C R_\uc/R},
\end{equation}
due to the collimation of the field lines, with $C \approx 3$ within
factors of order unity, and $R_\uc$ the poloidal radius of curvature
of the field lines.  In our approximation with $\nu\approx 0$, we have
$R_\uc \gg R$ so that $\gamma_2$ does not significantly change
$\gamma_\uff$.  In principle, $R_c$ can be computed as if the
field followed the shape of the disc, and then $B_r$ would be computed
by how the local magnetic flux changes in $\theta$.  However, these
effects are higher-order in tilt angle.

If the MHD jet is more conical (so has $\nu\sim 0$), the Lorentz
factor behaves qualitatively differently with a Lorentz factor given
by $\gamma = \gamma_\uff$, replacing the second asymptotic by the
solution to the cubic equation
\begin{equation}
\gamma_{2,\nu = 0} = C_1 \left[ \frac{(\mu - \gamma_{2,\nu = 0})}{\sin^2(\theta_c)} \ln{\left( 1 + C_2 \frac{r \Omega_\uF \sin(\theta_c)}{c \gamma_c} \right)} \right]^{1/3},
\label{gamma2m}
\end{equation}
where $C_1\approx 2$, $C_2 \approx 0.4$, $\sin(\theta_c) \sim
\cos(\theta) \approx 1$ for $\theta$ away from the equator, and $\gamma_c
\approx [\mu / \sin^2(\theta_c)]^{1/3}\approx (\mu)^{1/3}$
\citep{2012MNRAS.419..573M}.  Because in this paper we are interested
in $\nu \approx 0$, we set $\gamma_2 = \gamma_{2,\nu = 0}$.  This forces
the Lorentz factor to grow logarithmically with radius.  This gives
reasonable values of Lorentz factor as compared to GRMHD simulations
of discs and jets that have $\gamma \sim 5$ by $r \sim 10^3~r_\ug$
\citep{2006MNRAS.368.1561M,2009MNRAS.394L.126M}, e.g. for $\mu \sim
6$--$10$ at $r \sim 10^3~r_\ug$--$10^5~r_\ug$, this gives
$\gamma_{2,\nu = 0}\sim 4$--$8$.

\subsection{Torque equations}
\label{subsec:torque_equations}

For the given definitions of the internal and external torques, the
parallel and perpendicular torques can be written as
\begin{IEEEeqnarray}{rCl}
\tau_{\utot \parallel} & = &  - \tau_\uEM \sin(\theta_\udisc) + \tau_\uacc \cos(\theta), \\
\tau_{\utot \perp} & = & \tau_\uLT + \tau_\uEM \cos(\theta_\udisc) + \tau_\uacc \sin(\theta) ,
\end{IEEEeqnarray}
which are used in Eq.~(\ref{vectauperppar}).  As discussed above, to
lowest-order in $\theta$, the torque equation involving $\tau_{\utot
  \perp}$ controls the evolution of $\theta(r)$ and slight changes in
direction of the LT or BZ jet torque give the same result.

\section{Accretion Systems}
\label{sec:discsystems}
In this section, we discuss the types of disc systems we consider.

\subsection{Magnetic Field}

For disc models not based upon simulations, we use that model's
arguments for equipartition magnetic fields within the disc itself.
To get the field on the BH that drives the jet, we assume the radial
field strength in the disc is
\begin{equation}
B_r(r = r_\uH) \sim \left( \frac{H}{R} \right) |B|(r = r_\uH),
\label{BrfromB}
\end{equation}
as expected when small-scale turbulent eddies of size $H$ in the disc
set the scale of the radial field \citep{2001ApJ...548L...9M}.  This
leads to a radial field much weaker for thin discs than if one assumed
a dynamo existed that drove the toroidal field into a comparably
strong poloidal field.

GRMHD simulations show that magnetic flux can build-up to a saturated
value leading to the so-called magnetically-arrested disc (MAD).  MAD
simulations show that the BH magnetic flux magnitude is weakly
dependent on BH spin but depends linearly on disc thickness as:
\begin{equation}
\Upsilon \approx 10 \left( \frac{H/R}{0.3} \right),
\end{equation}
\citep{2011MNRAS.418L..79T,2012MNRAS.423.3083M,2015arXiv150805323A}.
Thin or thick non-MAD discs whose poloidal field is generated
spontaneously reaches up to $\Upsilon \sim 3$
\citep{2008ApJ...687L..25S,2010MNRAS.408..752P,2012MNRAS.423.3083M},
which is typically stronger than that obtained from nominal
equipartition arguments due to magnetic compression and lower plasma
$\beta_{\uMHD} \sim 1$ (i.e. gas to magnetic pressure ratio of order
unity) near the BH horizon
\citep{2010MNRAS.408..752P,2012MNRAS.423.3083M}.

\subsection{Advection-Dominated Accretion Flow for the limits of Radiatively Inefficient Accretion Flow and Super-Eddington Accretion}
\label{subsec:ADAF}

At low accretion rates, the flow is too tenuous to radiate efficiently
and becomes a RIAF. A classic model of a RIAF is the
Advection-Dominated Accretion Flow (ADAF) with most of the internal
energy being advected with the flow into the black hole before it can
be radiated away or before it can be blown into a wind
\citep{1994ApJ...428L..13N,1995ApJ...444..231N}.  A typical RIAF disc,
such as in \sgra{} or M87, may extend out to $10^4~r_\ug$ or $10^6~r_\ug$.

The extreme ADAF case with ratio of specific heats $\Gamma = 5/3$ has
purely radial accretion not typical of what is found in simulations.
To more closely connect with simulation results, we assume $\Gamma =
1.63$, corresponding to $\beta_\uMHD = P_\ugas/P_\umag = 5$
\citep{1999ApJ...516..399Q}. Furthermore we will assume the efficiency
of advection $f = 1$ (no radiative losses), and consequently
$\epsilon' = (5 / 3 - \Gamma) / [f (\Gamma - 1)] \approx 0.17$, $c_1 =
0.059$, $c_2 = 0.15$, $c_3 = 0.39$, which are the ratio of the radial
velocity to the Keplerian rotational velocity, the ratio of the
angular velocity to the Keplerian angular velocity, and the square
root of the ratio of the sound speed to the Keplerian rotational
velocity, respectively. Furthermore, $H / R = \sqrt{c_3} \approx 0.62$, and $\beta_0
= c_2 = 0.15$.  The surface density is given by:
\begin{equation}
\bar{\Sigma} = \left[ \frac{G^2 M_\odot}{c^4} m \right] \frac{\dot{M}}{2 \pi r v_r} = \frac{G}{c^3} \frac{\dot{M}_{\uEdd \odot}}{2 \pi} c_1^{-1} m \dot{m}_\uBH \bar{r}^{s - 1/2}.
\end{equation}
An ADAF wind can lead to less sub-Keplerian motion, for example
$s = 0.4$ gives $\beta_0\approx 0.53$ \citep{1999ApJ...520..298Q}.  In
order to estimate the magnetic field strength for the equipartition
state near the BH for the jet magnetic field strength, we assume
$\beta_\uMHD = 1$ seen in GRMHD simulations, which gives $\beta_\um =
P_\ugas / (P_\ugas + P_\umag) = 0.5$ for $P_\urad = 0$. To calculate the
dimensionless magnetic flux on the black hole, we assume that the
magnetic field, $B = \sqrt{2 \dot{M}_{\uEdd \odot}} \frac{c^{5/2}}{G
  M_\odot} m^{-1/2} \dot{m}^{1/2} \left( 1 - \beta_\um \right)^{1/2}
\alpha^{-1/2} c_1^{-1/2} c_3^{1/4} \bar{r}^{3/4}$ gives the total
field, and then we use Eq.~(\ref{BrfromB}) to obtain the radial
component.  This leads to a magnetic flux of $\Phi = 2 \pi r_\uH^2
|B_r|(r_\uH)$, and:
\begin{equation}
\Upsilon = \left( \frac{2 \pi}{\dot{M}_{\uEdd \odot}} \right)^{1/2}
\frac{G M_\odot}{c^{5/2}} m^{1/2} \dot{m}^{-1/2} r_\uH^2 B_r(r_\uH) .
\end{equation}
Otherwise, for a MAD RIAF, we use the simulation scaling of $\Upsilon$ as a function of $H/R$.

\subsection{Novikov \& Thorne disc}\label{nt}

At higher accretion rates, the disc becomes geometrically thin as
described by \citet{1973blho.conf..343N}, a general relativistic
extension to the disc model by \citet{1973A&A....24..337S}.  We
include the relativistic corrections and use the surface density of
the appropriate region based upon the opacity and gas or radiation
dominance as in the original paper by NT.  A typical BH X-ray binary
disc extends out to $10^5~r_\ug$ or $10^6~r_\ug$
(e.g. \citealt{2009ApJ...697..573O}).

To obtain the BH jet field strength for the equipartition magnetic
field case, we calculate the maximum magnetic field strength at twice
the innermost stable circular orbit (ISCO, which lies at $5~r_\ug$ for
$j = 0.9$) and assume the field at this location is the same total
strength as the strength over the horizon and then we use
Eq.~(\ref{BrfromB}) to obtain the radial component.  For the MAD state
we use $H / R\approx 0.257L/L_\uEdd$ corresponding to the maximum
thickness near the BH that would trap the hole's magnetic flux.  So,
\begin{equation}
\Upsilon = 10 \left( \frac{0.257 L / L_\uEdd}{0.3} \right) = 10 \left( \frac{0.257 \dot{m}}{0.3} \right)
\end{equation}
Although it is possible to modify the Novikov \& Thorne disc with a
wind \citep{2015ApJ...805...83M}, we have set $s = 0$ for this model.

\subsection{Slim disc}\label{slim}

In the high accretion rate limit, a reasonable analytical solution is
the slim disc model, which at high enough rates just becomes the ADAF
model for a $\Gamma = 4/3$ gas -- which is what we assume as our model
in this limit as valid in the limit of a radiation-dominated
flow. This gives $\epsilon' = 1$, $c_1 = 0.43$, $c_2 = 0.53$, $c_3 =
0.29$, and $H/R = \sqrt{c_3} \approx 0.53$, and $\beta_0 = c_2 =
0.53$. The smaller $H/R$ ratio causes a reduced (electro-)magnetic
torque, while the higher rotational velocity will increase the
accretion and Lense-Thirring torque, when compared with the low
luminosity case.  A typical tidal disruption event (TDE) disc might
extend out to $10^3~r_\ug$ or $10^6~r_\ug$
\citep{2015ApJ...812L..39D}.  We use the same magnetic field
conditions for the jet magnetic field strength as the ADAF case.

\subsection{GRMHD simulations}\label{grmhdsim}

For a general radiatively inefficient accretion flow (RIAF) we use the
results of numerical simulations. GRMHD simulations have a density
profile $\rho \propto \bar{r}^{-0.7}$ and $v_r \sim \alpha (H/R)^2
v_\phi$ for some total constant $\alpha (H/R)^2$
\citep{2004ApJ...611..977M,2010MNRAS.408..752P,2012MNRAS.423.3083M},
giving a surface density profile:
\begin{equation}
\bar{\Sigma} = \left[ \frac{G^2 M_\odot}{c^4} m \right] 2 \rho H = \frac{G}{c^3} \frac{\dot{M}_{\uEdd \odot}}{2 \pi} \bar{r}_+^{-2} m \dot{m}_\uBH \bar{r}^{0.3} .
\end{equation}
Note that one cannot infer the outflow parameter $s$ in
$\dot{M} \propto r^s$ from this density scaling, e.g. convective
dominated accretion flows (CDAFs) have shallow density profiles
compared to ADAFs even without unbound outflows.  However, this is not
required as we only need the density and $H$ scaling to get our
required $\Sigma$.  In practice, GRMHD simulations do not reach very
large radii, but such density scalings are seen in non-general
relativistic simulations that extend to large radii where a power-law
fit is quite reliable
\citep{2003ApJ...596L.207P,2011MNRAS.415.1228P}.

The $\phi$-velocity scaling seen in GRMHD simulations is quite
Keplerian for non-MAD discs, while for MAD simulations a slightly
steeper or shallower scaling may occur \citep{2012MNRAS.423.3083M},
which could slightly change our results.  For the non-MAD case, we use
$\beta_0 = 1$, while for the MAD state we take $\beta_0 = 0.5$ as
valid for $H/R \approx 0.4$ \citep{2012MNRAS.423.3083M} that is
applicable to RIAFs according to such simulations.

The BH jet magnetic field is set by the simulation condition of
$\Upsilon \sim 3$ for equipartition and set using the simulation
scaling of $\Upsilon$ with $H/R$ for a MAD.

\subsection{Overall Disc Model}

We branch between different disc solutions based upon $\dot{m}$ given
the estimated matching between disc theory and observations.  For
$\dot{m} = L / L_\uEdd < 10^{-3}$ we assume the system to be in a low
accretion RIAF state, using the ADAF state as detailed in section
\ref{subsec:ADAF} or the GRMHD simulation state as detailed in
section~\ref{grmhdsim}.  For $10^{-3} < \dot{m} < 10^{-2}$ we linearly
interpolate the relevant disc variables ($\beta_0$, $\beta_R$,
$\bar{\Sigma}$, and $\Upsilon_\uBH$), between the low accretion RIAF
and a Novikov \& Thorne disc described in section~\ref{nt}. For $10^{-2} <
\dot{m} < 1$, we use the Novikov \& Thorne solution only. For $1
< \dot{m} < 30$ we linearly interpolate between the Novikov \& Thorne
disc and the high accretion slim disc (in RIAF limit) as described in
Section \ref{slim}. For $30 < \dot{m}$ we use the high accretion slim
disc solution.

\section{Results}
\label{sec:results}

We assume the following values in all default cases, unless otherwise
specified: $\alpha = 0.1$, a moderately high spin of $j = 0.9$, $s =
0$, $\mu = 3$ for stellar-mass black holes and $\mu = 10$ for
supermassive black holes, $\theta_\uout = 5^\circ$, and
$r_\uout=10^5~r_\ug$.  By default, we assume the RIAF is modelled by
GRMHD simulation results.

\begin{figure*}
\begin{center}
\includegraphics[width = 0.46 \textwidth]{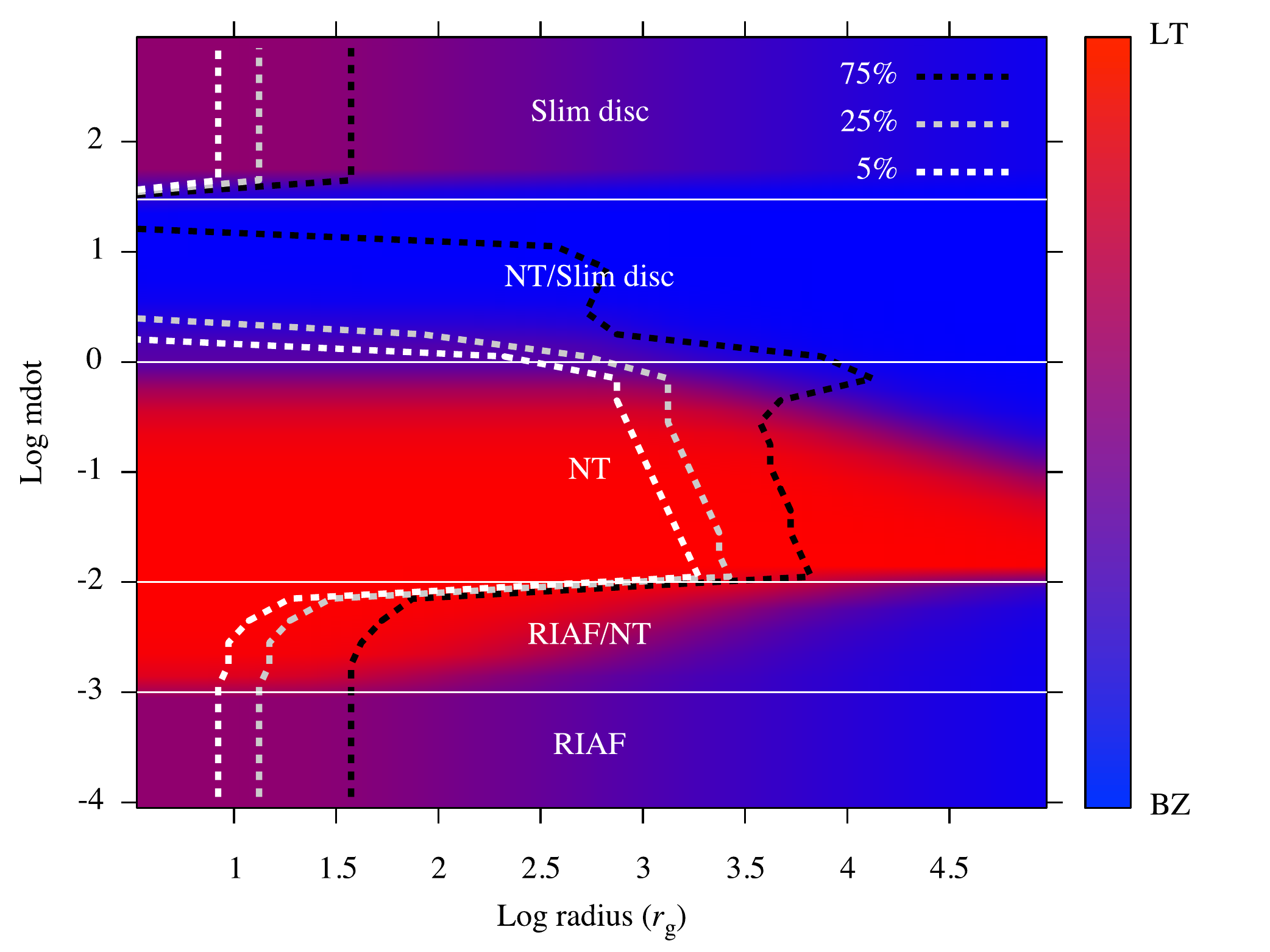}
\includegraphics[width = 0.46 \textwidth]{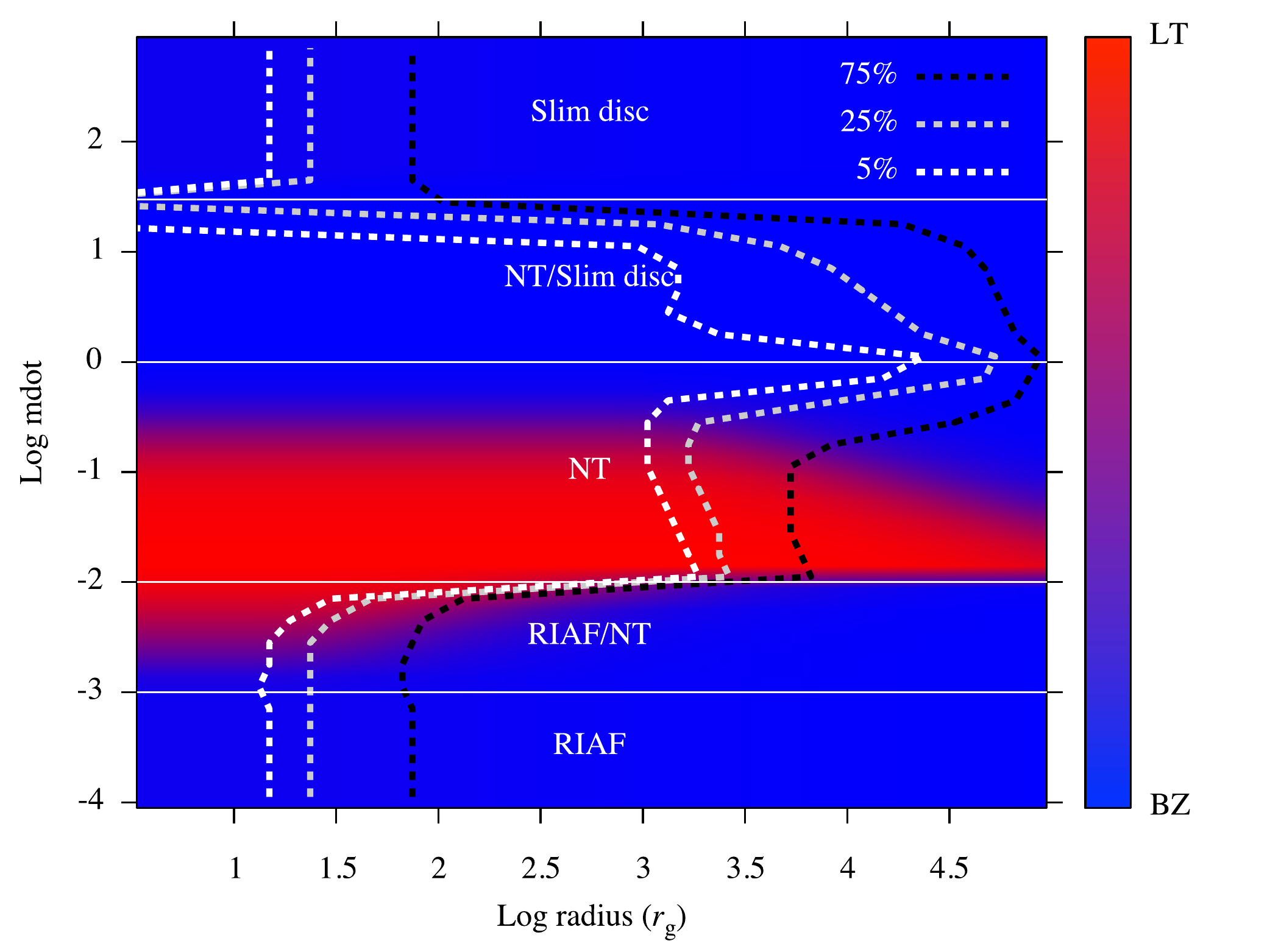}
\end{center}
\caption{Overview of the dominant torque for a range of radii and
  accretion rates for our overall disc model for the equipartition
  magnetic field model (left panel) and for the MAD magnetic field
  model (right panel) for \grs{}.  Blue corresponds to dominance
  by the BZ torque, while red corresponds to dominance by LT
  precession torques, with blends of colour in-between.  The rapid
  transitions at some $\dot{m}$ are due to the interpolation between
  the RIAF and NT discs.  This is caused by the surface density
  dropping monotonically with radius in a RIAF, while it rises quickly
  in the NT case. The dashed lines show when the angle of the disc has
  reached 75\%, 25\%, and 5\% of the original angle.  Our results are
  independent of the initial angle for small angles.  The NT disc is
  aligned effectively by LT torques at lower mass accretion rates,
  except if the field strength reaches MAD levels where the NT disc is
  aligned by the BZ torques before the LT torques can operate.  The
  RIAF models are only aligned with $r \sim 10~r_\ug$ due to the rapid
  infall time compared to the time it takes to modify the angular
  momentum by the LT or BZ torques.}
\label{fig:torque}
\end{figure*}

Fig.~\ref{fig:torque} shows our results for a broad set of mass
accretion rates for a black hole mass of $m=15$ consistent with
\grs{} (a typical stellar-mass black hole) in the non-MAD and MAD
states.  Alignment is effective (either by the LT or BZ jet torque) in
the NT state because of the relatively slow radial velocity for thin
discs.  This gives the LT or BZ jet torque time to align the disc as
material accretes inwards.  Naturally, the LT torque is only clearly
dominant for the thin disc case where the magnetic field is weak due
to the drop in magnetic flux as $H/R$ drops.  However, even in the NT
state the thickness is not too small for $\dot{m} \gtrsim 0.3$, beyond
which the BZ jet torque dominates alignment.  In either case, the NT
disc is aligned out to $r \sim 10^3~r_\ug$.

Alignment is not so effective in the RIAF state assuming the GRMHD
simulation scaling of density, but still alignment occurs within $r
\sim 10~r_\ug$ as primarily due to the radial component of the
magnetic field.  This is despite the relatively larger value of
$\Upsilon$, compared to the NT disc case, due to the thick disc at
$H/R\approx 0.4$.  Even in the equipartition magnetic field case, the
BZ jet torque tends to dominate the torques, but alignment is
dominated by LT torques -- so precession might occur instead for the
equipartition magnetic field case.

\begin{figure*}
\begin{center}
\includegraphics[width = 0.46 \textwidth]{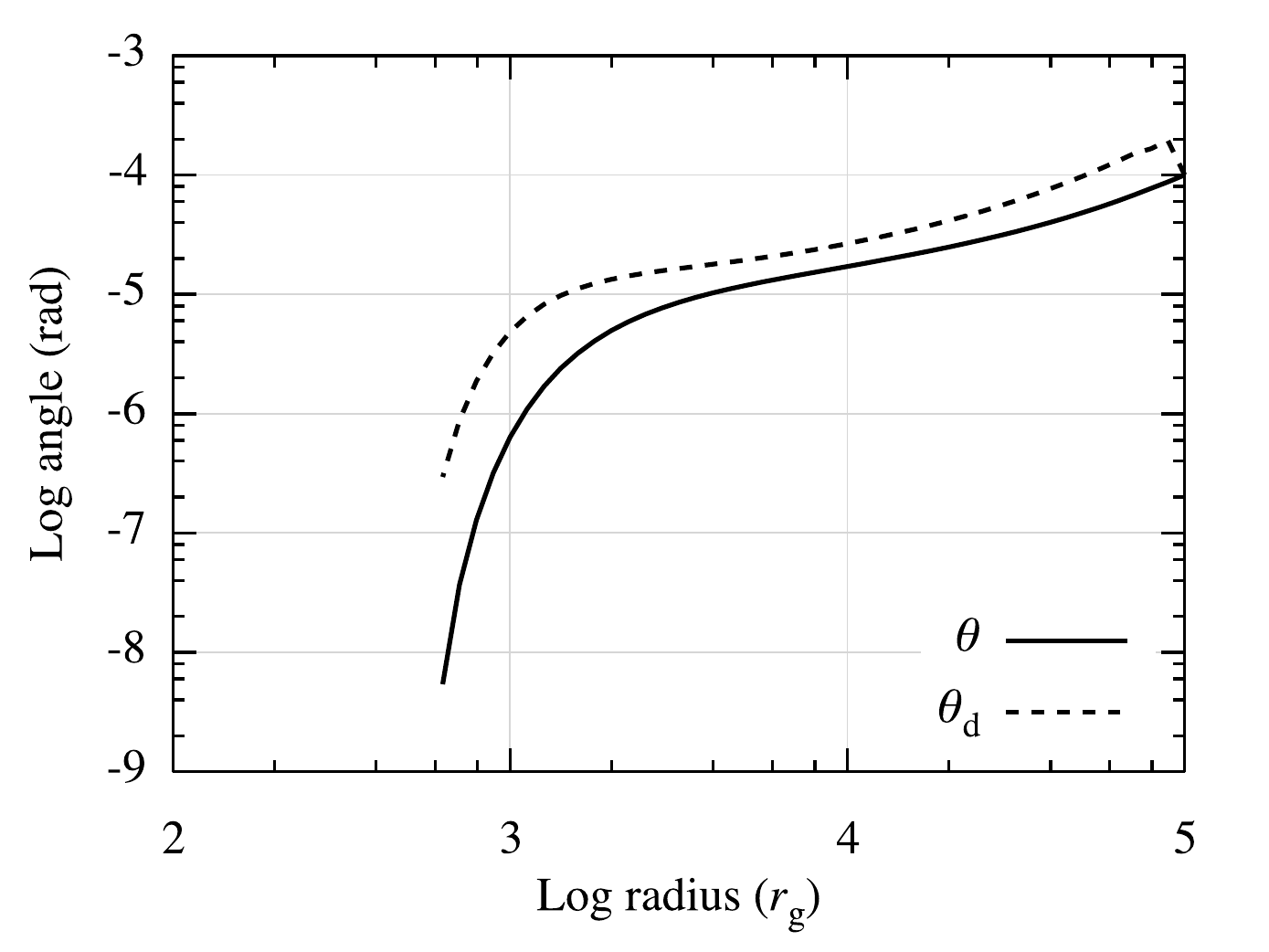}
\includegraphics[width = 0.46 \textwidth]{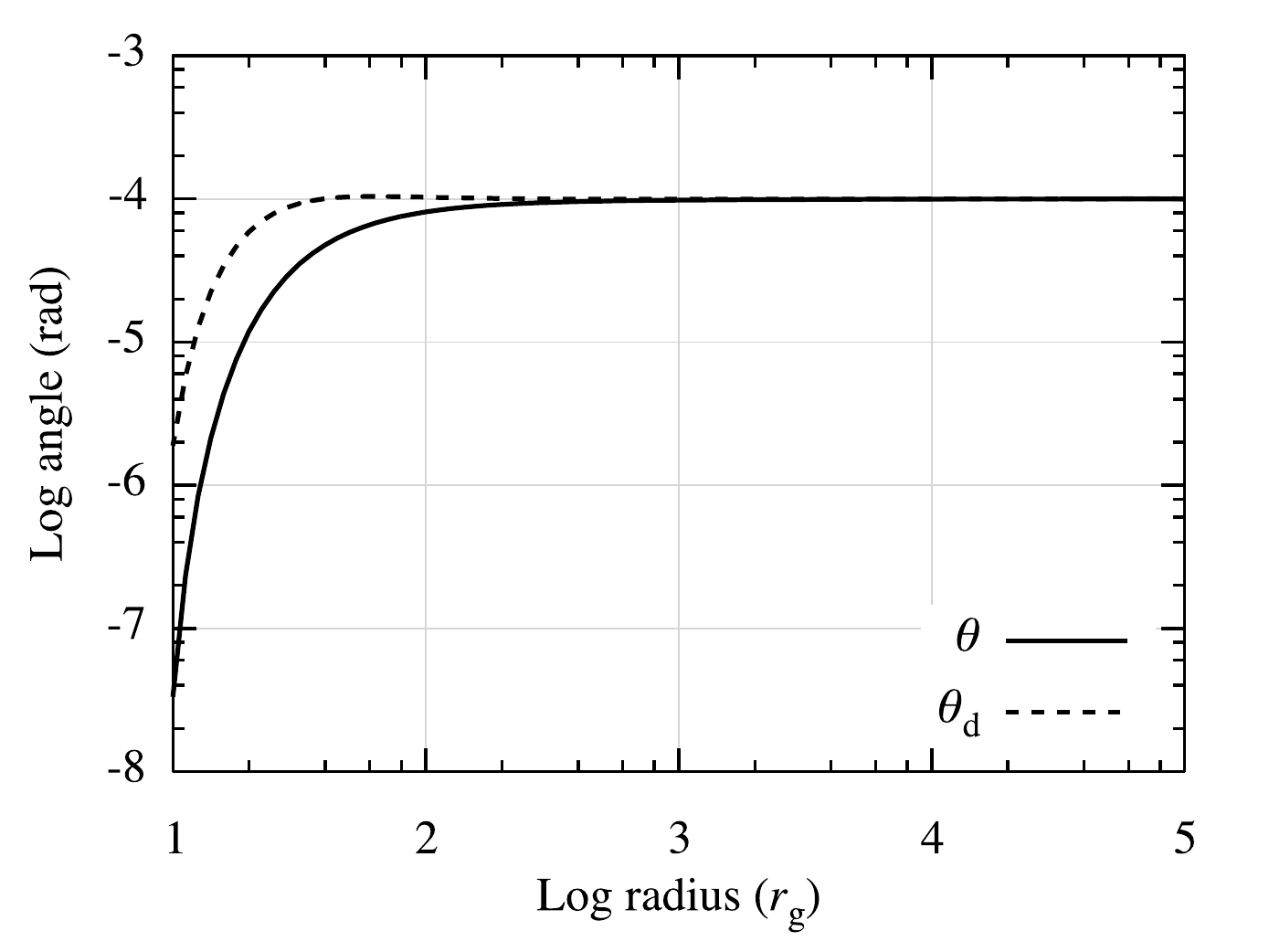}
\end{center}
\caption{Solution from torque equations for disc position angle
  $\theta$ and disc orientation angle $\theta_\udisc$ for the case of
  \grs{} in the high-soft state where $L / L_\uEdd
  = \dot{M} / \dot{M}_\uEdd \sim 0.3$ (left panel) and \sgra{}
  (right panel).  We have chosen the MAD RIAF simulation case for
  \sgra{} and chosen the MAD NT case for \grs{}.  In both cases,
  the disc extends out to $r_\uout\sim 10^5~r_\ug$, so we have chosen
  that as our outer disc radius.  In the standard thin disc case, at
  high enough $\dot{m} \gtrsim 0.3$, the BZ jet torque dominates the LT
  torque and causes alignment out to $r \sim 10^3~r_\ug$.  For the case of
  \sgra{}, the alignment occurs much closer to the BH at $r \sim 10~r_\ug$
  due to the rapid infall time.}
\label{fig:torquespecific}
\end{figure*}

Fig.~\ref{fig:torquespecific} shows the values of $\theta$ and
$\theta_\udisc$ for the case of \sgra{} with $m \approx 4.5 \times
10^6$, $\dot{m} \sim 1.5 \times 10^{-6}$, and $\mu = 10$ and for the
case of \grs{} (in the high-soft state where $L/L_\uEdd \sim
0.3$) with $m \approx 15$, $\dot{m} \sim 0.3$, and $\mu = 3$.  In both
cases the magnetic field is assumed to be MAD.  As indicated in
Fig.~\ref{fig:torque} for $\dot{m} \lesssim 0.3$, close to the black
hole the LT torque dominates, but for $\dot{m} \gtrsim 0.3$ the BZ jet
torque dominates and causes alignment out to $r \sim 10^3~r_\ug$ before
the LT torque has a chance to operate.  In the case of \sgra{}, the
inflow time is more rapid than in the standard thin disc case, but the
disc still aligns by the BZ jet torque by $r \sim 10~r_\ug$ as consistent
with GRMHD simulations of tilted discs around rotating black holes
\citep{2013Sci...339...49M}.

\subsection{LT versus BZ Jet Torque in Causing Alignment}

In some cases, the LT torque can appear to dominate, but we have
assumed a maximally aligning LT torque to test if the BZ jet torque
still dominates alignment.  We can also check the case when the LT
torque is weak (e.g., due to the LT torque instead leading to
precession) to see if the BZ jet torque alone is sufficient to lead to
substantial alignment.  If so, then even in those regions dominated by
the LT torque, the BZ torque is sufficient regardless of the regime
the LT torque operates in (i.e. diffusive versus wave-like).

We repeat the above analysis, but this time turn off the LT torque
entirely.  For the MAD case, we find that the BZ jet torque indeed
leads to substantial alignment even in regions where the LT torque
could dominate.  For example, in the transition region between NT and
slim disc regimes, the BZ torque alone is sufficient to align material
out to thousands of gravitational radii.

This means that even if LT leads to precession once $H/R$ increases
compared to $\alpha$ as $\dot{m}$ rises towards
Eddington, the BZ torque can still align the disc out to large radii.
So, the BZ torque can align the disc before the LT torque would have a
chance to cause precession.  In some cases, the intermediate regime
occurs, where the BZ torque could lead to alignment, but at such radii
the LT torque is already strong, in which case the disc could be
partially precessing and partially aligning.

\subsection{Dependence upon ADAF vs. GRMHD Simulations as RIAF Model}

The RIAF model applies at very low or very high accretion rates.  In
our default model, we select the simulation results as most accurately
portraying this regime.  Here instead we consider the ADAF model as our RIAF
model.  We repeat the same analysis as for our default case, but
consider how the RIAF regime changes.

Due to the steep density profile in the ADAF case, we find that the
jet rapidly aligns the disc out near the starting radius.  This is
because there is effectively much less angular momentum at large radii
compared to the default model.  The ADAF case leads to most of the
mass reaching the BH, leading to a maximally effective jet that can
more readily affect the outer disc.  In real discs, such as for \sgra{},
one expects the disc to transition to the Bondi inflow that itself
will transition to the ambient medium outside the influence of the
BH's gravitational potential (e.g. \citealt{2013ApJ...767..105L}).

The ADAF RIAF case is the only case where the results are also
significantly affected by how we treat the jet as either force-free or
MHD.  In particular, we have only considered the electromagnetic
contribution to any pressure by the jet on the disc.  As
electromagnetic energy is converted to kinetic energy in an MHD jet,
the electromagnetic contribution drops as our MHD jet case describes.
This leads to an ineffective electromagnetic pressure at large radii
in the ADAF RIAF case.  However, as described above, the total
pressure is best modelled by the force-free case, so that
the entire ADAF aligns out to large radius.

\subsection{Dependence upon other model parameters}

Here we consider the dependence on other model parameters, including
the outer disc radius $r_\uout$, total energy flux per unit rest-mass
flux in the jet $\mu$, dimensionless BH spin $j$, rotation rate
$\beta_0$, wind mass loss power-law index $s$, viscosity
parameter $\alpha$, and magnetic collimation parameter $\nu$.  In each case,
we repeat the analysis for new parameters.

For discs starting at smaller radii down to $r = 100~r_\ug$, the LT torque
plays a stronger role even for the RIAF regime.  This suggests that for
small discs, like those that may be present during some deep penetrating tidal
disruption events, the disc may tend to precess more than align.

For $\mu = 10$ (instead of our default $\mu = 6$) or slightly higher
values of $\mu = 20$, the results are similar with the expected less
effective magnetic pressure that scales as $1 / \gamma^2$ where
$\gamma \to \mu$ at large radii.  For smaller discs the change in $\mu$
leads to less dramatic effects because $\gamma$ has not reached terminal
values near $\mu$.

A smaller BH spin $j$ leads to a weaker jet as expected because the
magnetic pressure roughly scales as $j^2$.  Because the LT torque
scales as only $j$, this means eventually at small enough $j$ the LT
torque dominates (whether or not alignment occurs) and the BZ torque
plays no role as expected.

The rotation rate factor $\beta_0$ is typically fixed for each disc model
type, but smaller values lead to weaker LT torques and allow the BZ
torque to more easily dominate and lead to alignment because the disc
has less angular momentum for the same jet magnetic pressure (that
does not directly depend upon the disc rotation rate).  MADs tend to
be more sub-Keplerian, so the stronger field leads to a stronger jet
that even more easily aligns the more slowly-rotating material.

The mass-loss-rate power-law index $s$ that enters $\dot{M} \propto
r^s$ allows a shallower density profile, which makes the inner jet
less effective at aligning material present at fixed large radii.  The
simulation scalings used already effectively include a wind, and there
is still aligning at least within the region near the BH.

A smaller viscosity parameter of $\alpha=0.01$ leads to a slower
inflow rate, giving the LT torque or BZ jet torque more time to align
the material out to larger radii.  In this case, the BZ jet torque
aligns the RIAF simulation material out to $r \sim 100~r_\ug$ instead of
just $r \sim 10~r_\ug$.  The NT region is aligned by LT torques out to
$r \sim 10^4~r_\ug$ instead of just $r \sim 10^3~r_\ug$.

The magnetic collimation parameter $\nu$ is chosen as $\nu \approx 0$
as we do not include the disc pressure matching to the jet pressure.
Disc vertical stratification and pressure can force $\nu \approx 0.7$
\citep{2007MNRAS.375..513M}, leading to a stronger magnetic pressure
for any given radius.  This would lead to a much stronger BZ torque,
however a compensating factor is that collimating jets also become
causally disconnected from the disc and itself
\citep{2015MNRAS.452.1089P}.  The balance of these effects is left for
future work.

\section{Discussion}
\label{sec:discussion}

The Lense-Thirring torque has a steep dependence with radius, so it can
more readily dominate the BZ torque at small radii.  At larger radii,
the BZ jet torque can readily dominate.  If the BZ torque leads to
substantial alignment at large radii, then the LT torque has no chance
to operate and cannot align or lead to precession.  Whether alignment
occurs depends upon the disc model: NT models tend to align by LT and
other models tend to align by the BZ jet torque.

The BZ torque depends on the magnetic flux, and since large scale
magnetic fields can be more easily supported by thick discs
\citep{2001ApJ...548L...9M,2003PASJ...55L..69N,2012MNRAS.423.3083M},
it is not surprising that for a thick disc these torques are more
effective than for thin discs. For this reason in the thin Novikov \&
Thorne disc, the Lense-Thirring and accretion torque dominate, but for
even higher accretion rates, which cause the disc to become thicker,
the (electro-)magnetic torque becomes more dominant.

The LT torque depends on the density distribution, which can change
its radial profile. In the case of the inner region of a Novikov \&
Thorne disc, this dependency gives it a slope as flat as the accretion
torque, making it the strongest throughout that region. When a density
distribution with a negative radial dependence is used, the
Lense-Thirring torque quickly becomes negligible, sometimes not even
dominating at all. If most accretion happens at (super-)Eddington
rates, the disc will be thick, and the Bardeen-Petterson effect will
cause precession instead of alignment, also decreasing the importance
of the Lense-Thirring torque. For these two reasons the
Bardeen-Petterson effect can become negligible in these systems, a
hypothesis supported by recent numerical simulations \citep[and
  references
  therein]{2007ApJ...668..417F,2014ApJ...796..103M,2014ApJ...796..104Z}. As
a result we may need to look to other torquing mechanisms to align the
disc in astrophysical systems.

The Blandford-Znajek torque estimates in this paper assume a simple
monopolar field geometry ($\nu \approx 0$), while collimating jets
have a magnetic torque that drops off more slowly, and such geometries
may be more effective than we have assumed. On the other hand, the jet
can become ballistic before the Blandford-Znajek torque is completely
effective. These effects need to be modelled with detailed
simulations. We have included the equations for general values of
$\nu$ in this paper.

Our approach to computing the disc shape and including the torques is
quite simplified, but improves the one in \citet{1977A&A....58..175K}.
They compared the timescale of accretion to the timescale of magnetic
alignment for a disc that is rigid over an extended range of radii.
In our case, the disc shape is computed by the inclusion of advection
terms that account for the timescale of accretion versus alignment at
each radius.  We allow for stronger MAD-strength magnetic fields, and
we allow the disc to extend to physically reasonable larger radii
where the timescale for inflow becomes long enough for the BZ torque
to become more effective.

We have performed simple calculations of the disc shape due to
aligning torques caused by the BH spin axis being different than the
disc angular momentum axis. Our approach is rougher in handling the
Lense-Thirring torques than other analytical works
\citep{2011MNRAS.415.2122Z}, but no other work has considered the
competing effects of Lense-Thirring and Blandford-Znajek jet torques.
We treat the Blandford-Znajek torque fairly accurately.  Naturally,
prior analytical work is also unable to treat the magnetic field
within the disc accurately except as a viscosity.

Our calculation for the Lense-Thirring aligning torque is rough, but
we chose the torque to be maximally aligning with the goal of testing
whether the BZ jet torque could still dominate alignment.  Any other
missing physics that leads to precession, breaking, or oscillation by
Lense-Thirring would only weaken its ability to align before the BZ
jet aligns the disc.  In that sense, the calculation conservatively
determines where the BZ jet dominates alignment.

We have also taken a simple approach because simulations are expensive
and have yet to be self-consistent.  That is, simulations have
so-far been limited to using an effective viscosity instead of
turbulent magnetic fields
\citep{2000MNRAS.315..570N,2011MNRAS.415.2122Z}, tend to not resolve
the MRI and consider relatively thick discs in full GR
\citep{2014ApJ...796..103M}, or approximate GR with relatively thick
discs \citep{2013ApJ...777...21S}.  We have also neglected apsidal
precession, which could lead to additional external torques.  While
our calculations are rough, this has allowed us to scan a broader
parameter space of possible disc models, black hole masses, accretion
rates, and magnetic field types.  And it has allowed us to consider
the role of jet-induced alignment that otherwise requires expensive
simulations \citep{2013Sci...339...49M}.

The simulations performed by \citet{2013Sci...339...49M} show
alignment results reasonably consistent with our calculations, with
the RIAFs in the MAD state effectively aligning near the BH.  They
estimated the BZ jet torque analytically as an internal torque with
$\tau \sim r B_r B_\phi \sin(\theta)$ -- which is only roughly a good
approximation close to the black hole.

\section{Conclusions}\label{sec:conclude}

We have considered the competition between Lense-Thirring and
Blandford-Znajek jet torques in driving alignment between the disc
angular momentum vector and BH spin vector.  We have studied a broad
parameter space of possible disc models, black hole masses, accretion
rates, and magnetic field types in order to determine the disc shape,
what radius alignment occurs, and which torque dominates the alignment
process.

In those regions where BZ jet alignment is effective at aligning the
disc, Lense-Thirring becomes unimportant and ineffective because the
material is already aligned by the BZ jet before Lense-Thirring
operates.  This also means that Lense-Thirring will be ineffective at
other non-aligning processes like precession and disc breaking.

For BH X-ray binaries, state transitions show a variety of temporal
and spectral features that are difficult to explain theoretically
\citep{2006csxs.book..157M,2006ARA&A..44...49R,2006ApJ...652..518M}.
\grs{} is particularly prolific by exhibiting an array of
complicated temporal features.  Some of these temporal features occur
when \grs{} is near the Eddington rate, which we find is where
the MAD state can cause the BZ jet torque to align the material, while
for slightly sub-Eddington rates the Lense-Thirring torque dominates
and could cause some precession or alignment.  A combination of
precession by LT, alignment by LT and BZ torques, and quasi-periodic
oscillations driven by the MAD state and a BZ jet
\citep{2012MNRAS.423.3083M}, and variations in the magnetic field
(leading to different alignment and precessing torques) might interact
to produce such a diverse set of temporal phenomena.

If accretion happens at (or above) Eddington rates, the cosmological
evolution of black hole mass and spin may not be determined by the
Bardeen-Petterson effect, but rather by (electro-)magnetic
processes. The Blandford-Znajek torque can be a significant factor in
a wide range of accretion rates relevant to cosmological large-scale
structure simulations, so it should be taken into consideration.

The Event Horizon Telescope (EHT) focuses on the strong-field gravity
regime within tens of gravitational radii, and polarisation is
capable of probing the nature of the magnetic field there
\citep{2012ApJ...755..133S,2012MNRAS.421.1517D,2015Sci...350.1242J}.
Warps in the disc or bends in the jet near such scales could be
revealed by changes in the polarised emission by different
cancellations in the polarisation when the jet bends.  Our results are
consistent with the self-consistent GRMHD simulations by
\citet{2013Sci...339...49M}, who found disc alignment occurs within
$r \sim 10~r_\ug$.  This is on the scale observed by the EHT, so one
expects the warping of the disc or jet to manifest in EHT
observations.  Warps, oscillations, or precession could create
observable timing features \citep{2013ApJ...774L..22S}.  If warps are
present out to large radii, this can affect Faraday rotation measures
from discs and jets \citep{2010ApJ...725..750B}.  High-energy emission
from jets, as observed by Fermi, could have signatures of disc or jet
warps \citep{2015arXiv151008860O}.

The primary assumption made in this work is the small-angle
approximation, but this has allowed us to investigate a broad space of
disc models and parameters.  The hope is this inspires self-consistent
simulations that account for the competing roles between the
Lense-Thirring and Blandford-Znajek jet torques when considering the
alignment or breaking of discs.  Simulations are now able to account
for vertical stratification, winds, magnetic fields, and radiation,
which removes the ambiguity of the transitions between disc models and
can account directly for these competing effects
\citep{2014MNRAS.439..503S,2014MNRAS.441.3177M,2015MNRAS.454L...6M}.
Simulation models include much more physics than we have included, and
can treat the regime where the magnetic field undergoes polarity
inversions \citep{2014MNRAS.440.2185D}, which might light up the jet
to reveal the warped jet and disc.  Simulations in the thin disc
regime can be expensive, but our results show that even a study at
moderate Eddington factors ($L/L_\uEdd \sim 0.3$) might show
interesting competition between these torques while being in an
astrophysically relevant regime.

\section*{Acknowledgments}

We particularly thank Ramesh Narayan for very useful discussions
regarding how the jet torque would apply external pressure on the
disc.  We also thank Alexander Tchekhovskoy and Danilo Morales
Teixeira for useful discussions.  We acknowledge NASA/NSF/TCAN
(NNX14AB46G), NSF/XSEDE/TACC (TG-PHY120005), and NASA/Pleiades
(SMD-14-5451).

%\bibliographystyle{mn2e}
%\bibliography{references}

\label{lastpage}

\end{document}